\documentclass[aip,jcp,preprint,amsmath,amssymb]{revtex4-2}
\usepackage{graphicx}
\usepackage[update,prepend]{epstopdf}
\newcommand{\vv}{\mathbf}

\usepackage{natmove}

\begin{document}

\title{Dynamics of equilibrium linked colloidal gels}
\author{Taejin Kwon}
\affiliation{McKetta Department of Chemical Engineering, University of Texas at Austin, Austin, Texas 78712, USA}

\author{Tanner A. Wilcoxson}
\affiliation{McKetta Department of Chemical Engineering, University of Texas at Austin, Austin, Texas 78712, USA}

\author{Delia J. Milliron}
\affiliation{McKetta Department of Chemical Engineering, University of Texas at Austin, Austin, Texas 78712, USA}

\author{Thomas M. Truskett}
\email{truskett@che.utexas.edu}
\affiliation{McKetta Department of Chemical Engineering, University of Texas at Austin, Austin, Texas 78712, USA}
\affiliation{Department of Physics, University of Texas at Austin, Austin, Texas 78712, USA}

\begin{abstract}
Colloids that attractively bond to only a few neighbors (e.g., patchy particles) can form equilibrium gels with distinctive dynamic properties that are stable in time. Here, we use a coarse-grained model to explore the dynamics of linked networks of patchy colloids whose average valence is macroscopically, rather than microscopically, constrained. Simulation results for the model show dynamic hallmarks of equilibrium gel formation and establish that the colloid-colloid bond persistence time controls the characteristic slow relaxation of the self-intermediate scattering function. The model features re-entrant network formation without phase separation as a function of linker concentration, centered at the stoichiometric ratio of linker ends to nanoparticle surface bonding sites. Departures from stoichiometry result in linker-starved or site-starved networks with reduced connectivity and shorter characteristic relaxation times with lower activation energies. Underlying the re-entrant trends, dynamic properties vary monotonically with the number of effective network bonds per colloid, a quantity that can be predicted using Wertheim's thermodynamic perturbation theory. These behaviors suggest macroscopic in situ strategies for tuning the dynamical response of colloidal networks.
\end{abstract}
\maketitle

Colloidal gels are commonly formed by quenching a dispersion with low volume fraction, rapidly increasing the strength of short-range interparticle attractions relative to the thermal energy scale. For deep quenches, a strong thermodynamic driving force for macroscopic phase separation into colloid-poor and colloid-rich fluid phases competes with glassy frustration and increasing structural relaxation times.\cite{manley2005glasslike,gao2015microdynamics} As the latter exceeds the time scale for observation, a dynamically arrested percolated colloidal network forms.\cite{whitaker2019colloidal,petekidis2021rheology} Structure and structure-dependent (e.g., rheological and optical) properties of these nonequilibrium gels depend on the colloidal interactions and the quench protocol, as does the gel aging process.\cite{guo2011gel,eberle2012dynamical,negi2014viscoelasticity,johnson2019influence}

Colloids with orientation-dependent interactions that allow bonding to a only a few neighbors (e.g., patchy particles) can alternatively form equilibrium gels.\cite{Bianchi:2006,sciortino2017equilibrium} The fluid-fluid demixing transition for particles with such a microscopic valence restriction occurs at low volume fractions, opening up paths on the phase diagram between the demixing and glass transitions to directly form an equilibrium gel from a single-phase fluid. Equilibrium gels are soft and viscoelastic, exhibiting distinctive, spatially uniform dynamics that are stable in time.\cite{lattuada2021spatially} The dynamic density correlations in an equilibrium gel decay via a characteristic two-step process.\cite{biffi2015equilibrium,bomboi2015equilibrium,lattuada2021spatially} The intermediate-time plateau between the two, which reflects solid-like vibrational motion, has a height that is sensitive to temperature, increasing upon cooling as bonds form to extend the colloid network. The slow, secondary relaxation process, characterizing viscous flow, shows an Arrhenius temperature dependence with an apparent activation energy that is proportional to the strength of an attractive bond.\cite{biffi2015equilibrium,bomboi2015equilibrium,lattuada2021spatially}

Scalable synthesis of colloids with precisely defined valence poses formidable technical challenges, indicating a need for simpler and more generalizable strategies for assembling equilibrium gels. Linked colloidal assemblies, which restrict average colloidal valence, offer an attractive alternative that has motivated both modeling~\cite{hurtado2007heterogeneous,Lindquist:2016,Howard:2019,lowensohn2019linker,howard2021effects,howard2021wertheim,braz2021phase,xia2020linker,gouveia2022percolation,singh2022linker} and experimental~\cite{michel2000percolation,filali2001robust,Dominguez2020,Song2020,kang2021colorimetric,green2021assembling,sherman2022plasmonic} studies. Examples include nanocrystals with functionalized ligands and complementary linkers as well as emulsions with bridging telechelic polymers. Molecular thermodynamic theory and computer simulation of coarse-grained models have recently established that reducing the number of linkers per colloid can change the phase diagram in a way analogous to reducing the valence of equilibrium gel-forming patchy colloids.~\cite{Lindquist:2016,Howard:2019,sherman2021colloidal} Further, a re-entrant percolation transition as a function of linker concentration was predicted for a model linked-colloidal network,\cite{Lindquist:2016}
mirroring the re-entrant gelation reported experimentally for colloidal dispersions with bridging species concentration.\cite{zhao2012macrogel,luo2015gelation,Cabezas2018} These results suggest the need for a more complete understanding of how the dynamic properties of linked colloidal gel-forming systems relate to linker concentration, bonding motifs, and network structure.   

Here, we use simulation to characterize the dynamics of the coarse-grained PolyPatch (Polymer linked, Patchy colloid) model\cite{Howard:2019,howard2021effects}, in which the ends of ditopic molecules link neighboring nanoparticles into a network by attractively bonding to patches on their surfaces. Our simulations find dynamic signatures of a re-entrant equilibrium gel as a function of linker concentration, centered at stoichiometric conditions where the number of linker ends matches the number of nanoparticle surface bonding sites. The characteristic slow relaxation time of the self-intermediate scattering function scales with the colloid-colloid bond persistence time of the networks.
As predicted from Wertheim theory,\cite{howard2021wertheim} deviating from stoichiometry further restricts nanoparticle valence and the number of links per colloid that extend the colloidal network, producing linker-starved or site-starved networks. These structural changes correlate with reduced relaxation time scales and activation energies. Together, the results of this study help clarify how linked colloidal gels relate to patchy particle networks and suggest strategies for tuning the former's properties via macroscopic control parameters, some of which may enable real-time, in situ dynamic control.        

\section{Methods}
\label{sec:model}
\subsection{Model}
We adopt the PolyPatch model, a coarse-grained representation of dispersed colloids with surfaces decorated by six patchy sites that can attractively bond to the end beads of bifunctional linker molecules.\cite{Howard:2019,howard2021effects,howard2021wertheim,singh2022linker} The linkers are treated as flexible, linear chains of eight beads with diameter $\sigma$ and mass $m$. Each colloid is modeled as a spherical particle with diameter $\sigma_c$ = 5$\sigma$ and mass $m_c$ = 125$m$; its six interaction sites of diameter $\sigma$ are rigidly fixed in an octahedral arrangement a distance $r\approx 3.12\sigma$ from the colloid center.

All particles in the PolyPatch model interact with the hard-sphere-like repulsions of a core-shifted Weeks-Chandler-Andersen potential\cite{Weeks:1971}, i.e., 
\begin{equation}
    \beta u_{\rm r}(r) = \begin{cases}
    4 \left[\left(\dfrac{\sigma}{r-\delta_{ij}} \right)^{12} - \left(\dfrac{\sigma}{r-\delta_{ij}} \right)^6 \right] + 1, & r \le \sigma_{ij}^* \\
    0, & r > \sigma_{ij}^*
    \end{cases},
    \label{eq:wca}
\end{equation}
where $\beta = (k_{\rm B} T)^{-1}$, $k_{\rm B}$ is the Boltzmann constant, $r$ is the center-to-center separation of particles of types $i$ and $j$, and $\delta_{ij} = (\sigma_i + \sigma_j)/2 - \sigma$ modifies the divergence to reflect the respective diameters, $\sigma_i$ and $\sigma_j$. This repulsive interaction is truncated at $\sigma_{ij}^* = 2^{1/6} \sigma + \delta_{ij}$.

Two chemically bonded monomers of a linker chain also interact by a finitely extensible nonlinear elastic potential \cite{Bishop:1979},
\begin{equation}
    u_{\rm s}(r) = \begin{cases}
    -\dfrac{k_{\rm s} r_{\rm s}^2}{2} \ln\left[1 - \left(\dfrac{r}{r_{\rm s}} \right)^2 \right] + 4 \left[\left(\dfrac{\sigma}{r} \right)^{12} - \left(\dfrac{\sigma}{r} \right)^6 \right] + 1, & r < 2^{\frac{1}{6}}\sigma \\
    -\dfrac{k_{\rm s} r_{\rm s}^2}{2} \ln\left[1 - \left(\dfrac{r}{r_{\rm s}} \right)^2 \right] & 2^{\frac{1}{6}}\sigma \le r < r_{\rm s} \\
    \infty, & r \ge r_{\rm s}
    \end{cases}.
\end{equation}
with $\beta k_{\rm s} = 30\,\sigma^{-2}$ and $r_{\rm s} = 1.5\,\sigma$. \cite{Grest:1986} 

Attractions between colloidal interaction sites and linker ends can be expressed 
\begin{equation}
    \beta u_{\rm b}(r) = \begin{cases}
    -\beta \varepsilon_b \exp\left[-\left(\dfrac{5r}{\sigma}\right)^2\right], & r \le \sigma/2 \\
    0, & r > \sigma/2
    \end{cases},
    \label{eq:ub}
\end{equation}
where $\varepsilon_b$ is the bond strength. The repulsions of Eq.~\ref{eq:wca} and the range of the attractive interaction in Eq.~\ref{eq:ub} together ensure that two colloid sites cannot bond to the same linker end and two linker ends cannot bond to the same colloid site. Experimentally, thermoreversible linker-site bonds have been realized using DNA,\cite{mirkin1996dna,alivisatos1996organization,nykypanchuk2008dna,xiong2009phase,biffi2013phase,rogers2015programming,jones2015programmable} dynamic covalent bonding,\cite{kang2021colorimetric} and a variety of other reversible and orthogonal chemistries are available for use in linker-mediated assembly.\cite{Rowan2002,Jin2013,Seifert2016,borsley2016dynamic,Dominguez2020}.

\subsection{Simulation}
Molecular dynamics simulations of the PolyPatch model in the canonical ensemble were carried out using \textsc{lammps} (22 Aug 2018) \cite{Plimpton:1995}. An integration time step of $0.001\,\tau$ was chosen, where $\tau = \sqrt{\beta m \sigma^2}$ is a characteristic time, and temperature was maintained using a Langevin thermostat with friction coefficient $0.1\,m/\tau$ applied to each linker bead and colloid. To study how linker concentration impacts static and dynamic properties, the simulations included $N_{\rm c}=1000$ colloids and a range of linker-to-colloid ratios $\Gamma \equiv N_{\rm l}/N_{\rm c}$ from 1 to 6. It's helpful to recast the latter as the ratio of linker ends to colloid sites $\Gamma^*=\Gamma/3$, which varies from 1/3 to 2. This range encompasses behavior of linker-starved and site-starved state compositions, i.e., conditions below and above the stoichiometric ratio of $\Gamma^*=1$, respectively. The dimensions of the periodically replicated cubic simulation cell of volume $V$ were chosen to investigate colloid packing fractions $\eta_c = N_c \pi \sigma_c^3/(6V)$ from 0.01 to 0.15.      

 For each choice of $\eta_c$ and $\Gamma^*$, configurations were thermalized for $0.1 \times 10^5\,\tau$ in the absence of linker-site attractions ($\beta\varepsilon_b = 0$). Next, $\varepsilon_b$ was increased (at fixed temperature) for $0.1 \times 10^5\,\tau$ using a linear ramp until $\varepsilon_b=10$. After repeating the thermalization protocol at $\varepsilon_b=10$, $\varepsilon_b$ was again increased using a linear ramp of $0.8 \times 10^{-3}\,\varepsilon_b/\tau$. At a given $\beta\varepsilon_b$, molecular dynamics simulations were run until the potential energy converged; the required run time varied in the range $1.5 \times 10^3\,\tau$ -- $0.1 \times 10^7\,\tau$ depending on $\beta \varepsilon_b$ and $\Gamma^*$ . Finally, production simulations were performed to characterize the microstructure and the dynamics, with typical run times in the range $0.2 \times 10^5\,\tau$ -- $0.2 \times 10^7\,\tau$ for the state points considered. 

The behavior of the static structure factor at near-zero wavevector was used to estimate state points where macroscopic phase separation occurs. The partial static structure factor of the colloids was computed as\cite{Hansen2013}
\begin{equation}
    S_{\rm cc}(\vv{k}) = \frac{1}{N_{\rm c}} \left\langle \sum_{j,k}^{N_{\rm c}} e^{-i \vv{k}\cdot(\vv{r}_j-\vv{r}_k)} \right\rangle.
\end{equation}
Here $\vv{k} = (2\pi/V^{1/3})\vv{n}$ is a wavevector in the cubic box of volume $V$, $\vv{n}$ is a vector of integers, and $\vv{r}_j$ is the position of the $j$th colloid. The angle brackets indicate an ensemble average, which was calculated as a temporal average.  Specifically, we averaged $S_{\rm cc}(k)$ for the 22 smallest nonzero wavevector magnitudes $k=|\vv{k}|$ and extrapolated to $k=0$ by fitting the data to a Lorentzian function\cite{Jadrich:2015},
\begin{equation}
    S_{\rm cc}(k) \approx \frac{S_{\rm cc}(0)}{1+(k\xi_{\rm cc})^2},
\end{equation}
where $\xi_{\rm cc}$ is a correlation length. As in previous work, we used $S_{\rm cc}(0) > 10$ as an approximate indicator of phase separation \cite{Zaccarelli:2005,Lindquist:2016,Howard:2019}.

To characterize the colloidal dynamics, we calculated the mean squared displacement of colloidal particles $\left\langle (\Delta r)^2(t) \right\rangle$) as follows, 
\begin{equation}
    \left\langle (\Delta r)^2(t) \right\rangle =  \frac{1}{N_{\rm c}}\left\langle \sum_{j}^{N_{\rm c}} \left[\vv{r}_j(t)-\vv{r}_j(0)\right]^2, \right\rangle.
\end{equation}
where $\vv{r}_j(t)$ is the position vector of the $j$th colloid at time $t$. We also computed the colloidal self-intermediate scattering function:
\begin{equation}
    F_s(k,t) = \frac{1}{N_{\rm c}} \left\langle \sum_{j}^{N_{\rm c}} e^{-i \vv{k}\cdot \left[\vv{r}_j(t)-\vv{r}_j(0)\right]} \right\rangle.
\end{equation}
For sufficiently strong bonding interaction $\beta \varepsilon_b$, $F_s(k^*,t)$ decayed via a two-step relaxation, where $k^*$ is the position of the static structure factor's primary peak. To estimate a characteristic structural relaxation time $\tau_s$ and the nonergodicity factor $A_s$, we fit the slow relaxation of the self-intermediate scattering to the stretched exponential function, $F_s(k^*,t)=A_s \exp[-(t/\tau_s)^{\beta_1}]$ (Fig.~S1).

We computed two characteristic times for the bond dynamics:  continuous bond lifetime $\tau_b$ and bond persistence time $\tau_c$.\cite{luzar2000resolving,pal2003dynamics,del2008network,miller2009dynamical,del2010microscopic} Here, a bond refers to a linker-mediated connection between two colloids. Continuous bond lifetime was obtained from analyzing the bond survival probability $S_b(t)$, which can be expressed as 
\begin{equation}
S_b(t)=\left \langle \Sigma_{i,j} s_{ij}(t_0,t)\right \rangle/\left \langle \Sigma_{i,j} s_{ij}(t_0,0)\right \rangle
\end{equation} 
Here, $i$ and $j$ are colloidal indices; $s_{ij}(t_0,t)=1$ if colloids $i$ and $j$ are continuously bonded in the time interval $[t_0,t_0+t]$, and $s_{ij}(t_0,t)=0$ otherwise. The bond survival probability data was fit with an exponential function: $S_b(t)=\exp[-(t/\tau_b)]$ to obtain an estimate for the bond lifetime $\tau_b$ (Fig.~S2). The more permissive bond persistence time was obtained from the bond correlation function $C_b(t)$: 
\begin{equation}
C_b(t)=\left \langle \Sigma_{i,j} h_{ij}(t_0)h_{ij}(t)\right \rangle/\left \langle \Sigma_{i,j} h_{ij}^2(t_0)\right \rangle 
\end{equation}
Here, $h_{ij}(t)=1$ if colloids $i$ and $j$ are bonded at given time $t$, and $h_{i,j}(t)=0$ if the colloids are not bonded at that time (i.e., intermittent bond breaking is allowed). The bond correlation function was fit by a stretched exponential function, $C_b(t)=\exp[-(t/\tau_c)^{\beta_2}]$, to obtain an estimate of bond persistence time (Fig.~S3). 

Results for all dynamic properties were averaged over five independent molecular dynamics trajectories at each state point. Bars plotted with the simulated data points in the figures indicate standard deviations.

\subsection{Thermodynamic perturbation theory}
\begin{figure}
    \centering
    \includegraphics[width=1\linewidth]{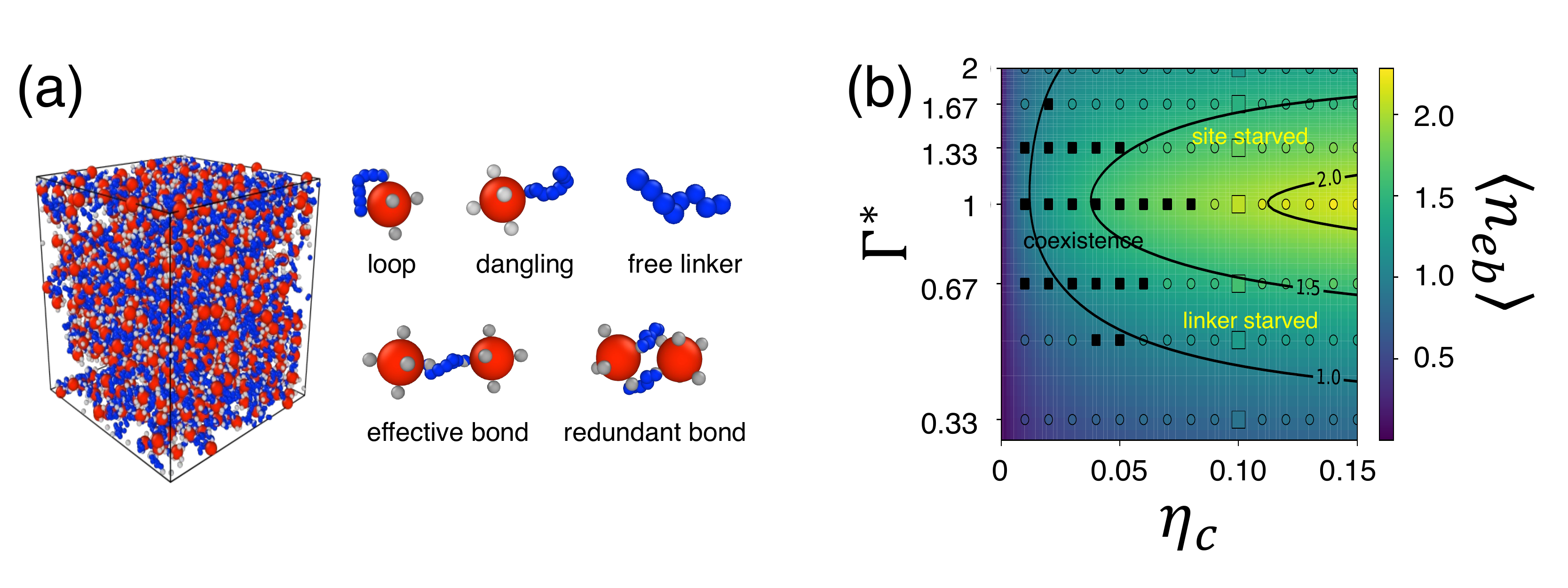}
    \caption{{\bf{Linker bond motifs and re-entrant network formation.}} (a, left) Simulation snapshot of the PolyPatch model. (a, right) Five linker bond motifs of linked colloidal networks. (b) Heat map of number of effective bonds per colloid $\langle {n}_{eb} \rangle$ in the $\Gamma^*-\eta_c$ plane for $\beta \varepsilon_b$=18. Open squares and circles indicate simulation data for $\langle {n}_{eb} \rangle$; their interior colors can be compared with the background colors and contour lines which reflect predictions of TPT1+ theory for the same. Open squares indicate $\Gamma^*$ values at $\eta_c = 0.10$ chosen for simulations of dynamic properties, conditions that span from linker-starved to site-starved states ($\Gamma^*=1$ is the stoichiometric ratio). Black squares represent simulation data with $S_{\rm cc}(0) > 10$, indicating the coexistence region where macroscopic separation between colloid-rich and colloid-poor phases occurs.} 
    \label{fig:phase}
\end{figure}

We use a recently introduced extension\cite{howard2021wertheim,howard2021effects} of Wertheim's first order thermodynamic perturbation theory\cite{Wertheim:1984a, Wertheim:1984b, Wertheim:1986a, Wertheim:1986b} (TPT1) to estimate the equilibrium linker bonding motifs of the PolyPatch model in the $\Gamma^*$--$\eta_c$ plane. We refer to this extension---distinctive in its ability to account for double-bonding motifs\cite{howard2021wertheim,howard2021effects}---as TPT1+. This theory can be solved to readily estimate the fractions of linkers that (1) form a \emph{loop}, (2) remain \emph{dangling} with a single end bonded to a colloid, (3) represent a \emph{free linker} with no ends bonded, (4) make an \emph{effective bond} between two colloids to extend the network, or (5) form a \emph{redundant bond} between two colloids (Fig.~\ref{fig:phase}a). A full description of the theory, including expressions for quantities needed for computing the number of each linker motif per colloid, can be found in Refs.~\citenum{howard2021effects} and \citenum{howard2021wertheim}.

Wertheim's perturbation theory decomposes the Helmholtz free energy density, $a$, into a sum of a bonding portion $a_{\text{bond}}$ and a reference contribution $a_0$. The former accounts for directional attractions between particles with interaction sites (i.e., linkers or colloids), while the latter accounts for the free energy of the system in the absence of such attractions. The bonding free energy density is expressed in terms of a series of variables, $X_\alpha^{(i)}$, each representing the fraction of component $i$  [i.e., colloid ($c$) or linker ($l$)] not bonded at a set of sites $\alpha$. Once the $X_\alpha^{(i)}$ are determined using TPT1+, one can compute the fraction of linkers in each of the five linker motifs.

In the parlance of TPT1+, if $A_{12} = \{A_1,A_2\}$ is the set of two (representative) neighboring bonding sites on a colloid and $B_{12} = \{B_1,B_2\}$ is the set of bonding sites on a linker, then the fraction of free linkers (i.e., with neither of two ends participating in a bond) is  $\chi_{\text{free}}=X_{B_{12}}^{(l)}$. Likewise, the fraction of dangling linkers (i.e., with either one, but precisely one, of its ends bonded) is $\chi_{\text{dangling}}=2(X_{B_{1}}^{(l)}-X_{B_{12}}^{(l)})$. We denote the fraction of linkers with both ends bonded
$\chi_{\text{both}}=1-\chi_{\text{dangling}}-\chi_{\text{free}}=1-2X_{B_{1}}^{(l)}+X_{B_{12}}^{(l)}$.

The fraction of linkers forming loops $\chi_{\text{loop}}$ and redundant bonds $\chi_{\text{redundant}}$ may be solved using relations formed in the derivation of TPT1+\cite{howard2021wertheim}
\begin{equation}
    \chi_{\text{loop}} = 2\nu_c\rho_cX_{A_{12}}^{(c)}X_{B_{12}}^{(l)}\Delta_2
\end{equation}
\begin{equation}
    \chi_{\text{redundant}} = 4\nu_c^2\left(\rho_cX_{A_{12}}^{(c)}\right)^2 \rho_l\left(X_{B_{12}}^{(l)}\right)^2\Delta_4
\end{equation}
where $\nu_c$ is the number of interaction site pairs on a colloid capable of participating in a loop or ring, $\rho_i$ is the number density of component $i$, and $\Delta_2$ and $\Delta_4$ are bond volumes introduced to account for loop and ring structures.\cite{howard2021effects}
The fraction of effective bonds $\chi_{\text{effective}}$ can be found with the relation
\begin{equation}
\chi_{\text{both}} = \chi_{\text{effective}} + \chi_{\text{loop}} + \chi_{\text{redundant}}.
\end{equation}
These fractions can be transformed from a per linker basis to a per colloid basis by multiplying by $\Gamma$. One quantity of particular interest is $\langle {n}_{eb} \rangle = \chi_{\text{effective}} \Gamma$, the number of effective bonds per colloid, which is related to a colloid's average coordination number $\langle CN_1 \rangle = 2 \langle {n}_{eb} \rangle$. $CN_1$ quantifies the number of colloids that a given colloid is directly bonded to via linkers.

\section{Results and Discussion}
\label{sec:results}
\subsection{Network formation and linker bonding}
Linker and colloid concentrations provide macroscopic handles for tuning network properties of linked-colloidal assemblies, as is illustrated by the behavior of the PolyPatch model. A heat map of $\langle {n}_{eb} \rangle$ in the $\Gamma^*$--$\eta_c$ plane at $\beta\varepsilon_b = 18$ predicted by TPT1+ shows a region of highly bonded network structures centered on the stochiometric ratio $\Gamma^*=1$ for $\eta_c$ of approximately 0.10 and higher (Fig.~\ref{fig:phase}b). The correspondence between the colors inside of the open symbols and the background reflects the excellent agreement between the theoretical predictions and simulated results for $\langle {n}_{eb} \rangle$. At lower $\eta_c$ and in the vicinity of $\Gamma^*=1$, effective bonding can be satisfied only by macroscopic phase separation, while at higher $\eta_c$ a single phase with sufficient bonding to percolate is found, consistent with the expectation of equilibrium gelation. Crossing contours of $\langle {n}_{eb} \rangle$ at constant $\eta_c$ corresponds to re-entrant network formation that is approximately symmetric in $\ln \Gamma^*$. Lower values of $\langle {n}_{eb} \rangle$, and consequently the colloidal coordination number, that attain for conditions away from the stoichiometric ratio highlight that colloidal valence becomes restricted at low $\Gamma^*$ due to linker starving (i.e., too few linkers per colloid), while it is limited at high $\Gamma^*$ due to a lack of unoccupied colloidal sites to form effective bonds (i.e., site starving).    

\begin{figure}
    \centering
    \includegraphics[width=1\linewidth]{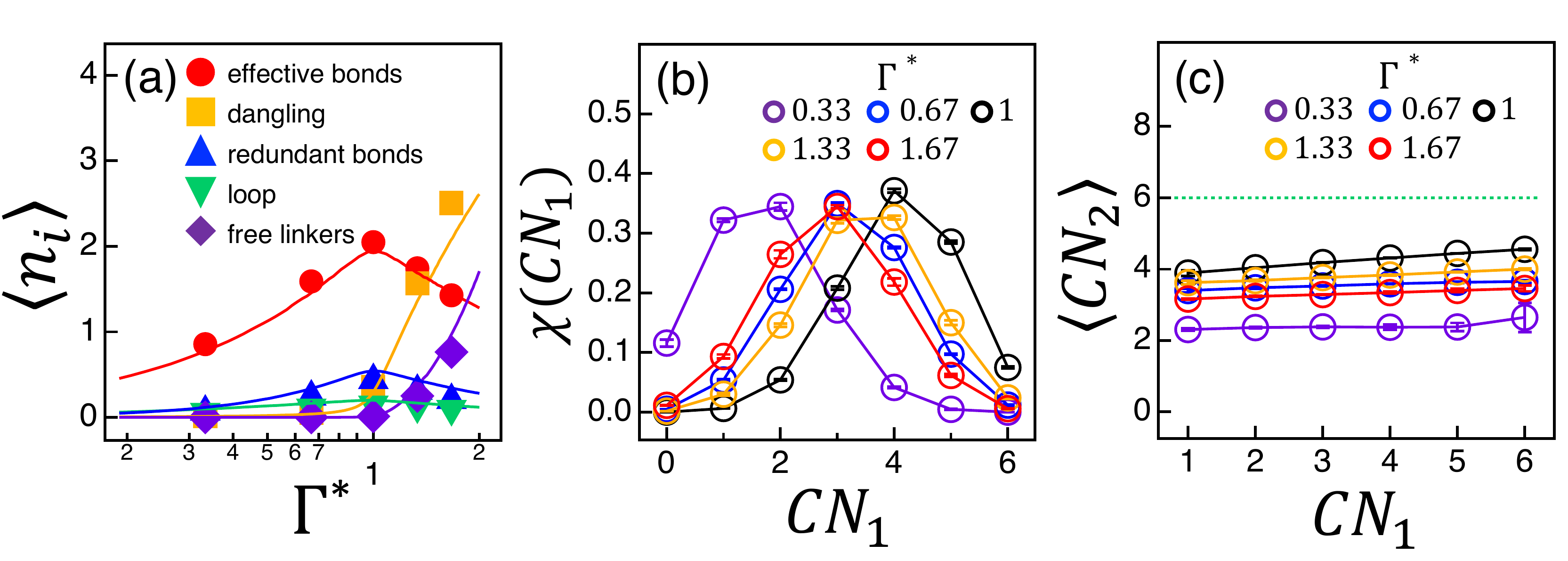}
    \caption{{\bf{Macroscopic tunability of network connectivity.}} (a) Number of linker bond motifs per colloid $\langle {n}_{i} \rangle$ versus $\Gamma^*$, where $i$ represents motif type from Fig.~\ref{fig:phase}a. Closed symbols represent simulation results and curves are predictions of TPT1+ theory. (b) Fraction of colloids $\chi$ with coordination number $CN_1$. (c) Coordination number $\langle CN_2 \rangle$ for colloids bonded to a central colloid with coordination number $CN_1$. Green dotted line represents maximum possible value of $CN_2$. Results in all panels are for $\beta \varepsilon_b=18$ and $\eta_c = 0.10$.}
    \label{fig:structure}
\end{figure}

To further probe the microstructural ramifications of changes to linker concentration, we computed the number of each of the five linker bonding states shown in Fig.~\ref{fig:phase}a per colloid as a function of $\Gamma^*$ at $\eta_c=0.10$ (larger squares in Fig.~\ref{fig:phase}b). The changes in the various bond motifs per colloid predicted by TPT1+ are in good quantitative agreement with those computed via molecular simulation (Fig.~\ref{fig:structure}a). In the linker starved region ($\Gamma^*<1$), there is no shortage of available colloidal interaction sites, and thus adding linkers increases the number of connections between colloids and extends the linked network. On the other hand, in the site starved region ($\Gamma^*>1$), adding linker chains decreases the number of linkers bridging colloids, displacing them to form dangling linkers.

In addition to these modifications to the linker bonding motifs, changing $\Gamma^*$ has a pronounced effect on the colloidal coordination number $CN_1$. Histograms showing the simulated fractions of colloids with different coordination numbers, $\chi(CN_1)$, are unimodal and shift non-monotonically with $\Gamma^*$. This trend reflects re-entrant structural modifications as the system passes through the stochiometric ratio, where the colloids have the highest average coordination number (Fig.~\ref{fig:structure}b). The peak positions of the distributions closely track their average values $\langle CN_1 \rangle = 2 \langle {n}_{eb} \rangle$, further highlighting that $\langle {n}_{eb} \rangle$ captures key aspects of network connectivity.

To ascertain how much the bonding state of a colloid influences that of its neighbors, we computed $\langle CN_2 \rangle$, the average coordination number of colloids linked to a central colloid with $CN_1$ (Fig.~\ref{fig:structure}c). To a first approximation, $\langle CN_2 \rangle \approx \langle CN_1 \rangle$, again following the re-entrant trends of $\langle {n}_{eb} \rangle$  in Fig.~\ref{fig:structure}a  as $\Gamma^*$ is increased. For each $\Gamma^*$, there is a weak, positive correlation between $\langle CN_2 \rangle$ and $CN_1$; neighbors of a more highly coordinated particle are themselves bonded to slightly more neighbors than the neighbors of low-coordinated colloids. This positive correlation, reflecting network cooperativity, is strongest at the stoichiometric ratio and flattens for $\Gamma^*$ in linker-starved or site-starved regions. 

\subsection{Colloidal dynamics}
\begin{figure}
    \centering
    \includegraphics[width=0.8\linewidth]{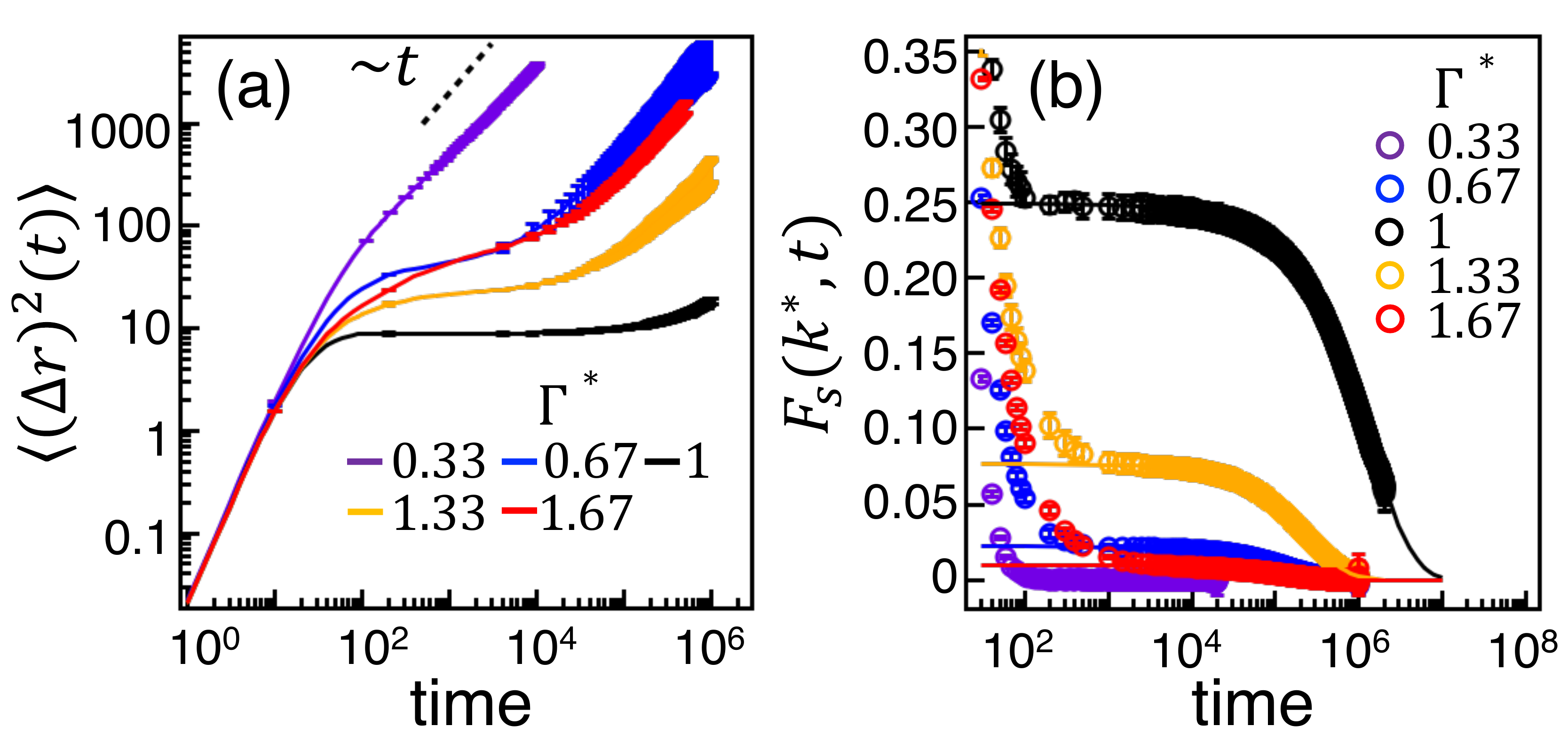}
    \caption{{\bf{Colloidal dynamics in re-entrant viscoelastic networks.}} (a) Mean squared displacements of colloids $\left\langle(\Delta r)^2\right\rangle$. (b) Self-intermediate scattering functions of colloids $F_s(k^*,t)$. Symbols represent simulation results and curves are stretched exponential function fits. All results are for $\beta \varepsilon_b =22$ and $\eta_c=0.10$.}
    \label{fig:dynamics}
\end{figure}

To study how linker concentration influences colloidal relaxation processes in the PolyPatch model, we computed the mean squared displacement $\langle (\Delta r)^2\rangle$ and the self-intermediate scattering function $F_s(k^*,t)$ of the particles, where $k^*\sigma=1$ corresponds to the approximate position of the colloidal structure factor's first peak (Fig.~\ref{fig:dynamics}). At $\beta \varepsilon_b=22$ and $\Gamma^*=0.33$, i.e., a strong-bonding but linker-starved condition, $\langle (\Delta r)^2\rangle$ shows equilibrium fluid-like behavior, crossing over from short-time dynamics to long-time diffusive scaling without exhibiting a plateau at intermediate times. The corresponding intermediate scattering function displays a fast, single-step decay. As $\Gamma^*$ is increased toward the stoichiometric ratio, plateaus form in $\langle (\Delta r)^2\rangle$ and $F_s(k^*,t)$ at intermediate times reflecting solid-like vibrational dynamics, while a second, viscous relaxation process becomes evident at long times. The height of the plateau of $F_s(k^*,t)$, commonly referred to as the nonergodicity parameter $A_s$, also shows pronounced increases as $\Gamma^*$ approaches one and the linked-colloidal network becomes more extensive. However, as even more linker is added, creating site-starved conditions, the viscoelastic characteristics of the single-particle dynamics attenuate, and the relaxation processes again approach simple fluid-like behavior.

\begin{figure}
    \centering
    \includegraphics[width=1\linewidth]{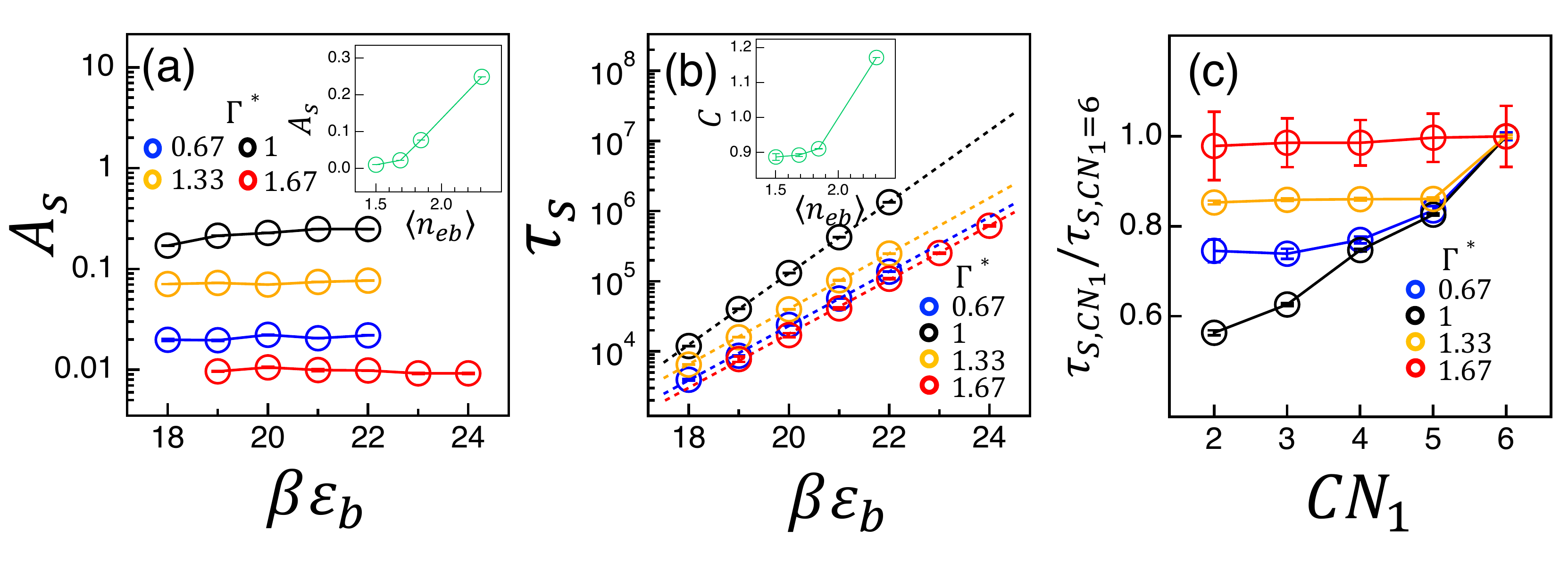}
    \caption{{\bf{Arrhenius colloidal dynamics at, and away from, the stochiometric ratio.}}
    (a) Nonergodicity parameter $A_s$ versus $\beta \varepsilon_b$ or (inset) number of effective bonds per colloid $\langle {n}_{eb} \rangle$. (b) Structural relaxation time $\tau_s$ versus $\beta \varepsilon_b$. Symbols represent simulation results and lines are Arrhenius fits, $\tau_s=\tau_0\exp(C\beta \varepsilon_b)$ where $\tau_0$ and $C$ are fitting parameters. (inset) $C$ versus number of effective bonds per colloid $\langle {n}_{eb} \rangle$. (c) Normalized structural relaxation time for subpopulations of colloids bonded to $CN_1$ neighboring colloids at $\beta \varepsilon_b=22$.  All results are for $\eta_c = 0.10$.}
    \label{fig:relaxation}
\end{figure}

Linker concentration and strength of attraction between linker ends and colloidal surface sites play distinctive roles in tuning dynamics of colloidal networks (Fig.~\ref{fig:relaxation}). At fixed $\beta \varepsilon_b$, changing $\Gamma^*$ from 0.67 to 1 increases both $A_s$ and the characteristic slow relaxation time $\tau_s$ by over an order of magnitude, dynamic effects that are reversed upon increasing $\Gamma^*$ further to 1.67 (Fig.~\ref{fig:relaxation}a and \ref{fig:relaxation}b). Underlying this re-entrant behavior is a monotonic change in $A_s$ with $\langle {n}_{eb} \rangle$ along this path; higher values of $A_s$ reflect more constrained vibrational motion of colloids in strongly interconnected networks. Varying $\beta \varepsilon_b$ at fixed $\Gamma^*$ reveals an Arrhenius relationship for the slow relaxation, $\tau_s=\tau_0\exp(C\beta \varepsilon_b)$, with activation energy proportional to the bond strength, i.e., $C\varepsilon_b$. These temperature-dependent dynamics are consistent with those observed in equilibrium gel-forming fluids of particles with microscopically restricted valence, e.g., DNA nanostars and patchy colloids\cite{rovigatti2011self,biffi2015equilibrium,roldan2017connectivity,russo2021physics}, as well as in strong glass-forming liquids\cite{russo2021physics}.

Similar to the nonergodicity parameter, the activation energy prefactor $C$ shows re-entrant behavior with $\Gamma^*$, assuming lower values ($C \approx 0.9$) in both the linker- and site-starved states and higher values ($C \approx 1.2$) at the stoichiometric ratio. These trends mirror those observed in patchy colloids and other model network-forming fluids.\cite{smallenburg2013liquids,roldan2017connectivity,shi2018common,russo2018water,rovigatti2018self,russo2021physics} Activation energies in the latter can vary from $0.5 \varepsilon$ to $2 \varepsilon$, with the larger values corresponding to more geometrically constrained networks where particle motion requires correlated, simultaneous breaking of multiple bonds.\cite{russo2021physics,de2006dynamics,nezbeda1989primitive,nezbeda1990primitive,roldan2017connectivity}

To gain further insights into how changes in $\Gamma^*$ lead to different dynamic behaviors, a self-intermediate scattering function analysis was also carried out at $\beta \varepsilon_b = 22$ for individual subpopulations of colloids with $CN_1=i$ with $i \in \{2,3,4,5,6\}$ at $t=0$. (Particles with $CN_1=1$ do not significantly contribute to two-step relaxation.) In addition to increasing the average coordination number of colloids and their neighbors (Fig.~\ref{fig:structure}b and \ref{fig:structure}c), tuning $\Gamma^*$ toward the stoichiometric ratio also produces coordination-number sensitive relaxation times (Fig.~\ref{fig:relaxation}c). Similar coordination-number dependent single-particle dynamics were reported for simulated tetrahedral network forming patchy particles.\cite{roldan2017connectivity}  In both the PolyPatch model at $\Gamma^*=1$ and the tetrahedral patchy particle fluid\cite{roldan2017connectivity}, bond breaking is essential for colloidal translation (Fig.~S4a).   
However, in linked-colloidal systems, modifying $\Gamma^*$ to linker-starved or site-starved conditions with fewer effective bonds per colloid substantially weakens the link between dynamics and coordination number (Fig.~\ref{fig:relaxation}c), and at $\Gamma^*=1.67$ there are no detectable differences in relaxation times of particles with different $CN_1$. The reduced connectivity of the colloidal network under these conditions renders it less constrained, and significant colloidal displacements can occur without even rupturing bonds with the neighbors (Fig.~S4b).

\subsection{Bond dynamics}

\begin{figure}
    \centering
    \includegraphics[width=1\linewidth]{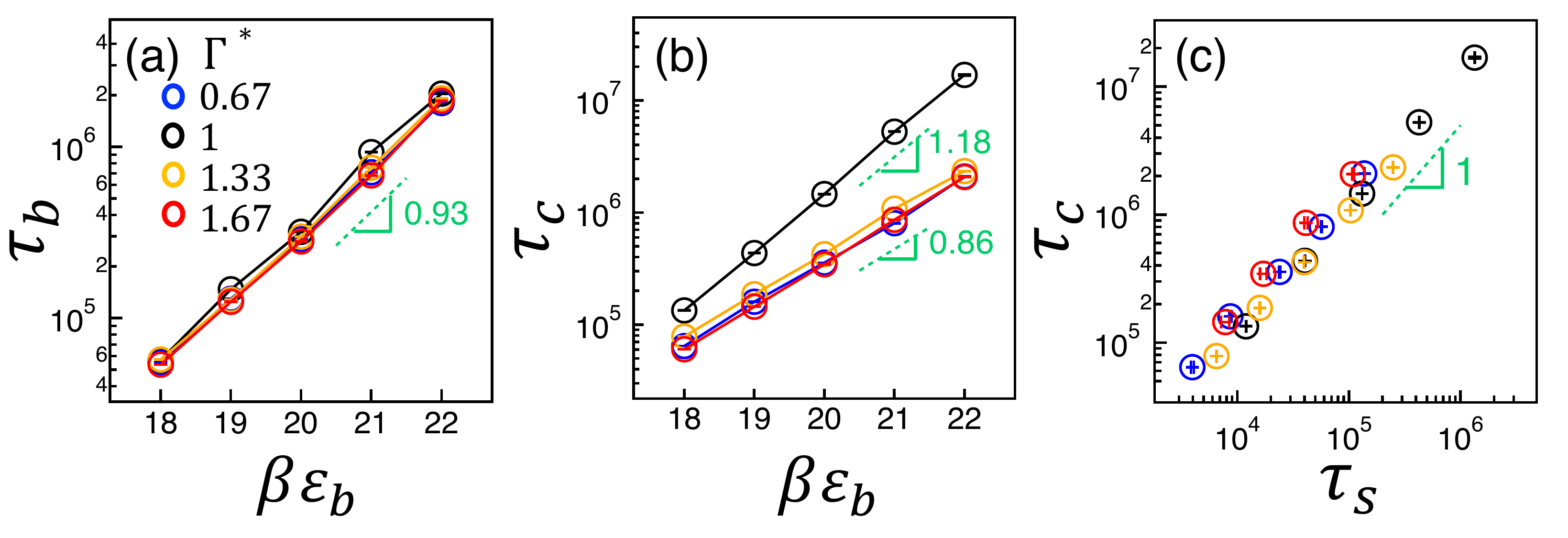}
    \caption{ {\bf{Colloid-colloid bond persistence controls structural relaxation}}
    (a) Continuous colloid-colloid bond lifetime $\tau_b$ versus $\beta \varepsilon_b$. (b) Colloid-colloid bond persistence time $\tau_c$ versus $\beta \varepsilon_b$. (c) Colloid-colloid bond persistence time $\tau_c$ versus structural relaxation time $\tau_s$. All simulated data are for $\eta_c = 0.10.$}
    \label{fig:bonddynamics}
\end{figure}

A comparison of continuous bond lifetime $\tau_b$ (i.e., time for a colloid-colloid bond to break), and bond persistence time $\tau_c$ (i.e., time that a colloid-colloid bond lasts allowing intermittent breaks) for the PolyPatch model provides insight about the microscopic processes underlying differences in structural dynamics as $\Gamma^*$ varies.  Both times are found to exhibit Arrhenius temperature dependencies for the full range of conditions examined in this study (Fig.~\ref{fig:bonddynamics}a and \ref{fig:bonddynamics}b). However, the former is found to depend only on $\beta \varepsilon_b$ and not on $\Gamma^*$, which determines the connectivity of the network and the constraints it imposes on the colloids. By contrast, the bond persistence time $\tau_c$ shows a $\Gamma^*$-dependent activation energy, which is larger at the stoichiometric ratio than under linker-starved or site-starved conditions. This makes physical sense; at a given $\beta \varepsilon_b$, looser, less interconnected networks allow colloids to diffuse away from a bonding partner after a linker end detaches from a colloidal site, while more geometrically constrained networks tend to promote bond reformation between the same pair. A direct comparison of data in Fig.~\ref{fig:relaxation}b and \ref{fig:bonddynamics}b  shows $\tau_c \propto \tau_s$ (Fig.~\ref{fig:bonddynamics}c), highlighting that $\Gamma^*$-dependent persistence of a colloid-colloid bond (allowing for intermittent bond breaks), controls the slow relaxation of the self-intermediate scattering function in these viscoelastic networks.


\section{Conclusions}
\label{sec:conclude}
By studying a coarse-grained model, we explored how linked colloidal networks can exhibit dynamic hallmarks\cite{biffi2015equilibrium,bomboi2015equilibrium,lattuada2021spatially} of equilibrium-gel forming colloids with microscopically restricted valence. These dynamic properties include a characteristic two-step relaxation of the self-intermediate scattering function with a nonergodicity parameter that increases as gelation is approached within a single fluid phase (i.e., avoiding phase separation) and a structural relaxation time with an Arrhenius temperature dependency. Unlike patchy particles, linker concentration provides a macroscopic handle in linked-colloidal gels for tuning network connectivity and viscoelastic dynamics. Re-entrant properties are observed as linker concentration is tuned away from the stoichiometric ratio where the number of linker ends matches the number of colloidal sites. Decreasing linker concentration reduces the connectivity of the network by starving the system of linkers that connect sites on neighboring colloids.  Increasing linker concentration has a similar effect by displacing effective bonds connecting two colloids with dangling linkers. Through a bond lifetime analysis, we demonstrated that the colloid-colloid bond persistence time (with intermittent bond breaking) controls the characteristic slow relaxation of the self-intermediate scattering function.

Looking foward, our study suggests that equilibrium gel dynamics in linked-colloidal networks could similarly be tuned, but at constant linker and colloid concentration, by adding a capping species that binds to colloidal sites, displacing connecting linkers. In DNA nanostar gels, similar strategies have produced temperature-re-entrant networks.\cite{bomboi2016re} Synthetic methods for creating dynamically activated (e.g., light-triggered) caps may further allow in situ control of network connectivity and properties. Given the accuracy of TPT1+ for predicting the bonding motifs of the PolyPatch model, the theory should prove to be a powerful tool for predicting how the presence of caps influences the underlying equilibrium properties of such networks.

In future studies, it would be helpful to extend the PolyPatch model to more accurately treat the dynamic covalent bonding strategies that are being implemented in experimental linked-colloidal networks.\cite{Dominguez2020,kang2021colorimetric,green2021assembling,sherman2022plasmonic} Specifically, incorporating the thermodynamics and kinetics of the reactions that govern bonding between linker ends and colloidal surface sites (e.g., ligands) would allow the model to provide insights into how macroscopic rheological properties could be programmed using different dynamic covalent chemistries.\cite{fitzsimons2021effect,jackson2022designing}

\begin{acknowledgements}
This research was primarily supported by the National Science Foundation through the Center for Dynamics and Control of Materials: an NSF MRSEC under Cooperative Agreement No.~DMR-1720595, with additional support from the Welch Foundation (Grant Nos.~F-1696 and F-1848). We acknowledge the Texas Advanced Computing Center (TACC) at The University of Texas at Austin for providing HPC resources.
\end{acknowledgements}

\section*{Data Availability}
The data that support the findings of this study are available from the authors upon reasonable request.


%

\section*{Supplementary Information}

\renewcommand{\thefigure}{S1}

\begin{figure}
     \centering
     \includegraphics[width=1\linewidth]{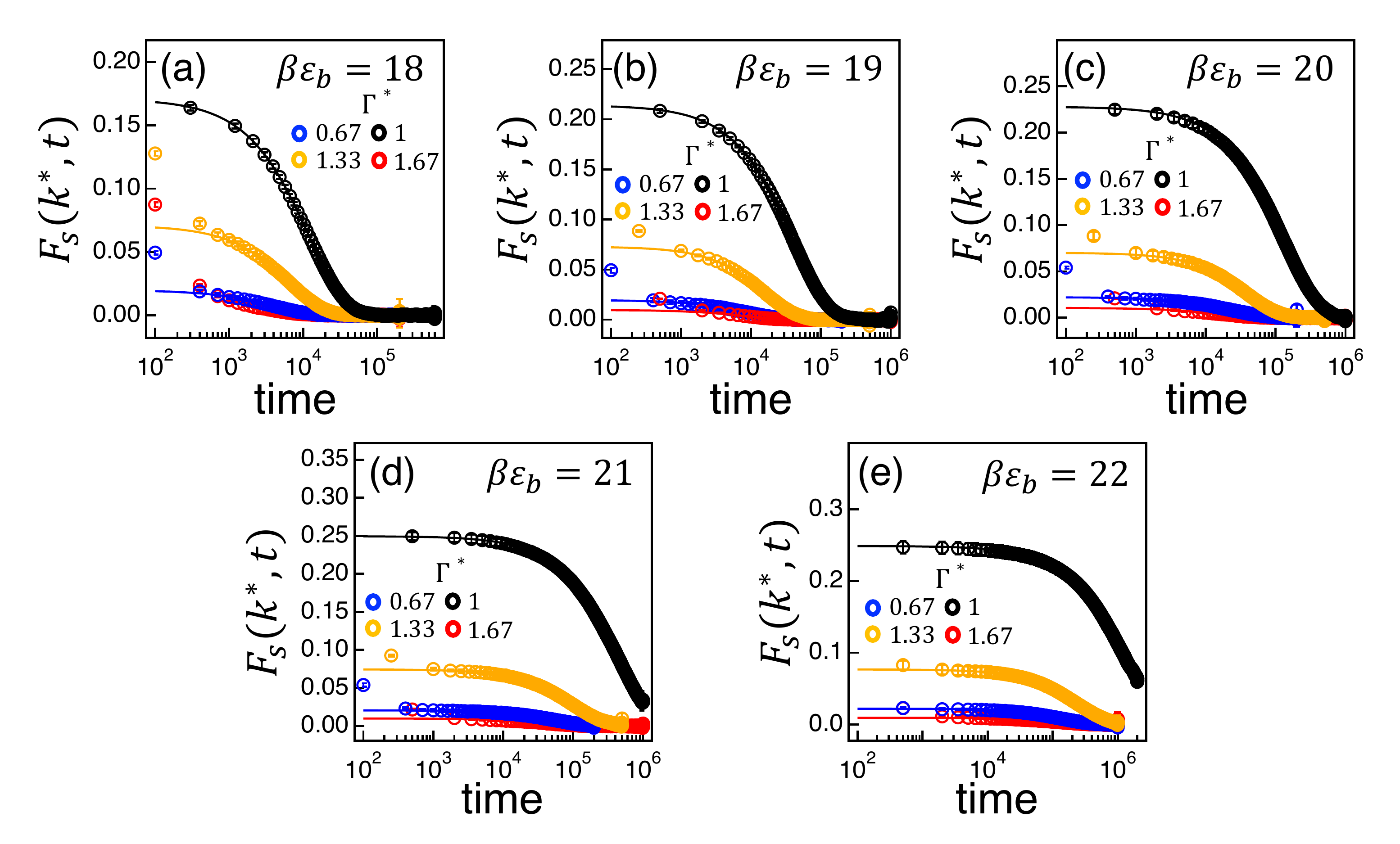}
     \caption{Self-intermediate scattering functions of colloids $F_s(k^*,t)$ for different $\Gamma^*$ at (a) $\beta \varepsilon_b$=18, (b) $\beta \varepsilon_b$=19, (c) $\beta \varepsilon_b$=20, (d) $\beta \varepsilon_b$=21, and (e) $\beta \varepsilon_b$=22. Symbols represent simulation results and curves are fitted stretched exponential functions, $F_s(k,t)=A_s \exp[-(t/\tau_s)^{\beta_1}]$, where $A_s$, $\tau_s$, and $\beta_1$ are fitting parameters.}
     \label{fig:fkt_si}
\end{figure}

\renewcommand{\thefigure}{S2}

\begin{figure}
     \centering
     \includegraphics[width=1\linewidth]{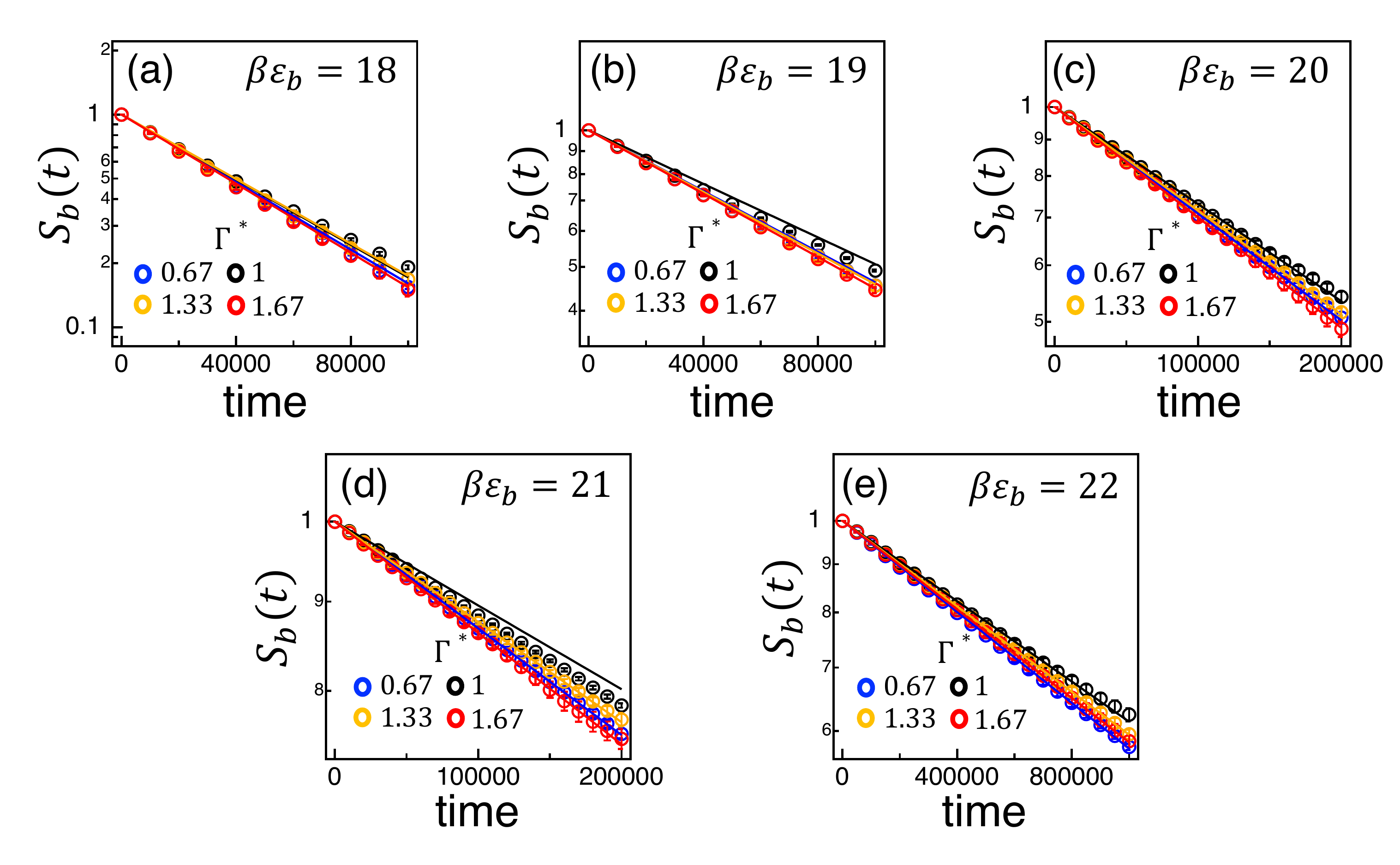}
     \caption{Bond survival probability $S_b(t)$ for different $\Gamma^*$ at (a) $\beta \varepsilon_b$=18, (b) $\beta \varepsilon_b$=19, (c) $\beta \varepsilon_b$=20, (d) $\beta \varepsilon_b$=21, and (e) $\beta \varepsilon_b$=22. Symbols represent simulation results and curves are fitted exponential functions, $S_b(t)=\exp(-t/\tau_b)$, where $\tau_s$ is a fitting parameter.}
     \label{fig:bondsurivival_si}
\end{figure}

\renewcommand{\thefigure}{S3}

\begin{figure}
     \centering
     \includegraphics[width=1\linewidth]{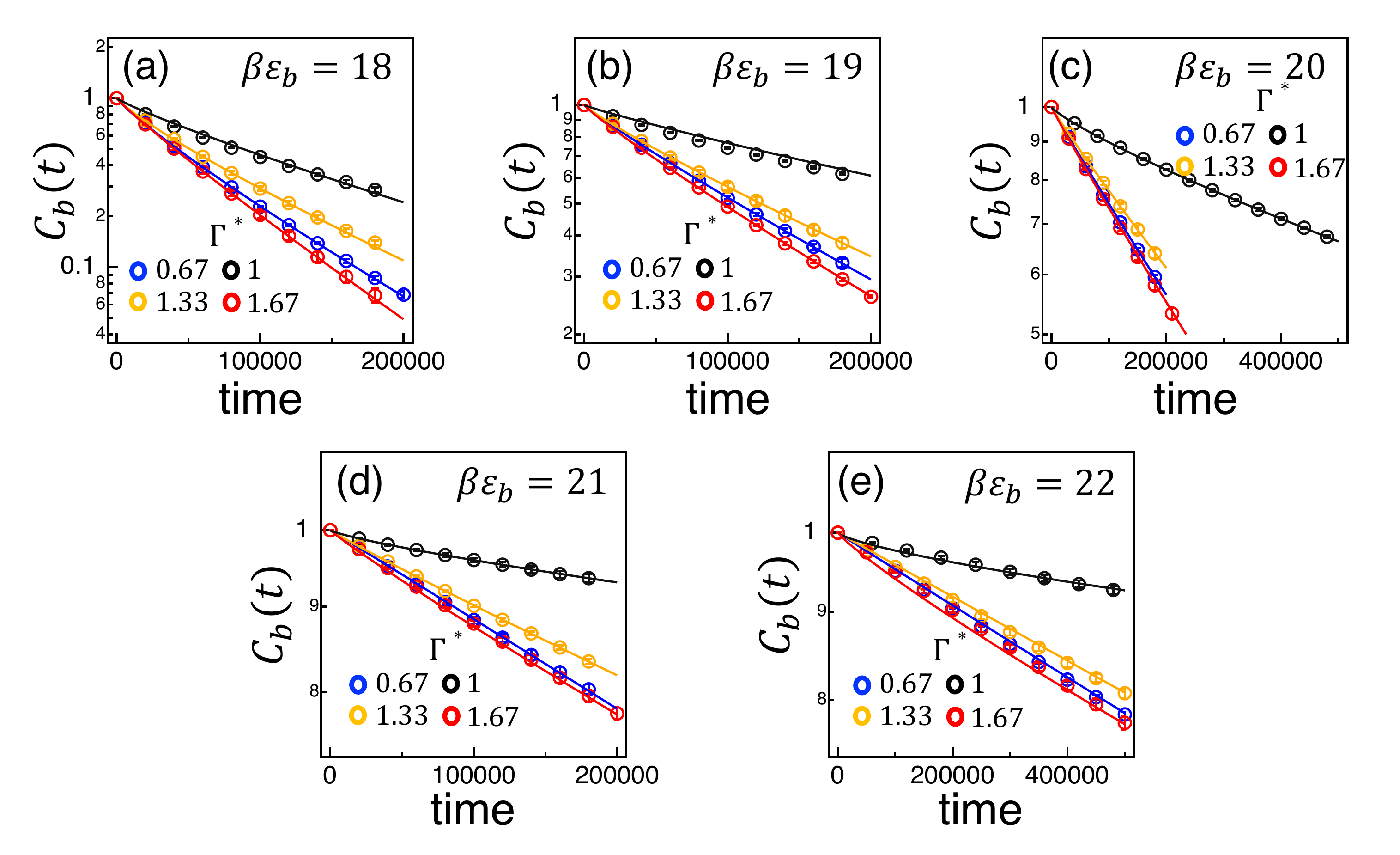}
     \caption{Colloidal bond correlation function ($C_b(t)$) for different $\Gamma^*$ at (a) $\beta \varepsilon_b$=18, (b) $\beta \varepsilon_b$=19, (c) $\beta \varepsilon_b$=20, (d) $\beta \varepsilon_b$=21, and (e) $\beta \varepsilon_b$=22. Symbols represent simulation results and curves are fitted stretched exponential functions, $C_b(t)=\exp[-(t/\tau_c)^{\beta_2}]$, where $\tau_c$ and $\beta_2$ are fitting parameters.}
     \label{fig:bondcorr_si}
\end{figure}

\renewcommand{\thefigure}{S4}

\begin{figure}
     \centering
     \includegraphics[width=1\linewidth]{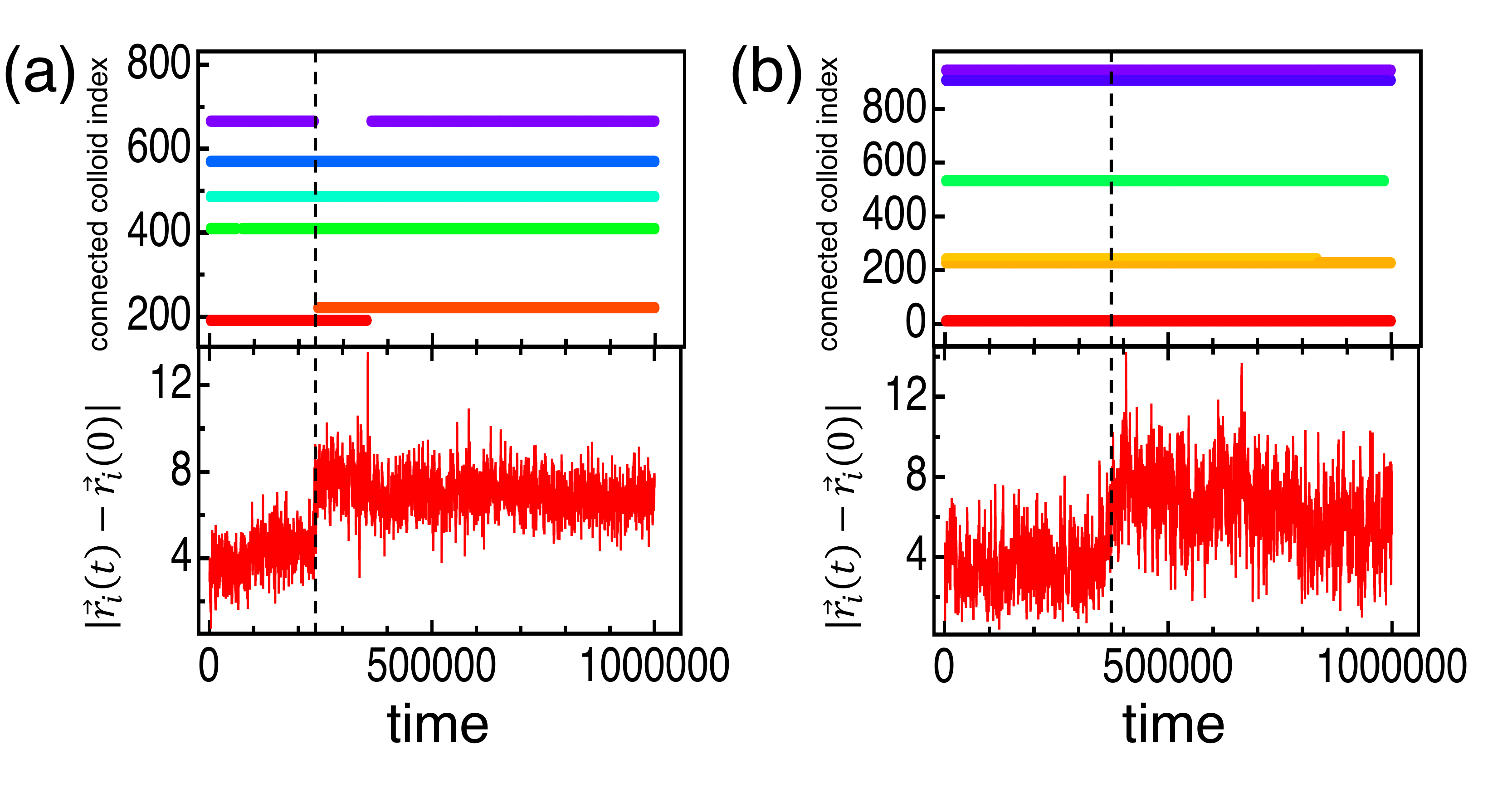}
     \caption{(top) Temporal profile for connected colloidal neighbors of a selected colloid and (bottom) the magnitude of displacement of a selected colloid from the initial position at $t$=0 at $\beta \varepsilon_b$=22, $\eta_c=0.10$, and (a) $\Gamma^*$=1 or (b) $\Gamma^*$=1.67. Colors in the upper figures represent neighboring colloid index. Black vertical dotted lines represent the large change of displacement of a selected colloid.}
     \label{fig:singledynamics_si}
\end{figure}

\clearpage

\end{document}